\documentclass{iau}

\usepackage{graphicx}

\title[Is There a Relation between Duration and $E_{\rm iso}$ in Gamma-Ray Bursts? ] 
{Is There a Relation between Duration and $E_{\rm iso}$ in Gamma-Ray Bursts? }

\author[ Shu-Jin Hou, Tong Liu, Da-Bin Lin, Xue-Feng Wu, \& Ju-Fu Lu]   
{Shu-Jin Hou$^1$$^,$$^2$,
  \ Tong Liu$^1$,
  \ Da-Bin Lin$^1$,
  \ Xue-Feng Wu$^2$
   and \ Ju-Fu Lu$^1$}

\affiliation{$^1$ Department of Physics and Institute of Theoretical Physics and Astrophysics,\\University Xiamen, Fujian 361005,China \\ email: {\ houshujingrb@163.com } \\[\affilskip]
$^2$ Purple Mountain Observatory,Chinese Academy of Sciences, Nanjing 210008,  China \\email: {\ xfwu@pmo.ac.cn}}

\pubyear{2012}
\volume{290}  
\jname{Feeding compact objects: Accretion on all scales}
\editors{C.M. Zhang, T. Belloni, M. M\'endez \& S.N. Zhang, eds.}

\begin{document}

\maketitle

\begin{abstract}
The system of accretion disk and black hole is usually considered as the central engine of Gamma-ray Bursts (GRBs).
It is usually thought that the disk in the central engine of GRBs is the advection-dominated accretion disk,
which is developed from a massive (mass $M_{\rm disk}$) torus at radius $r_{\rm disk}$.
We find a positive correlation between the isotropic gamma-ray energy $E_{\rm iso}$ and duration (so-called $T_{\rm 90}$)
for GRBs. We interpret this correlation within the advection-dominated accretion disk model, associating $E_{\rm iso}$ and $T_{\rm 90}$
with $M_{\rm disk}$ and viscous timescale respectively.
\keywords{gamma-ray: bursts; accretion, accretion disks; black hole physics}
\end{abstract}

\firstsection 
\section{Introduction}
GRBs are the most luminous events observed at cosmological distances.
The physics of GRBs remains as a great puzzle, especially that of the central engine.
The burst duration $T_{\rm 90}$ is generally regarded
as a mark of the activity duration of the central engine.
In general, the observed GRBs are classified into two groups by $T_{\rm 90}$ divided at 2 s (\cite[Kouveliotou et al. 1993]{Kouveliotou93}),
i.e., long and soft GRBs $vs$ short and hard GRBs, corresponding to two physical types of GRBs produced by deaths of massive stars and mergers of two compact objects, respectively.
According to current understanding, whether long or short bursts, the progenitors all form a hyperaccreting black hole. Here we adopt the popular torus-accretion model for GRBs (\cite[Popham et al. 1999]{Popham99}; \cite[Narayan et al. 2001]{Narayan01}; \cite[Liu et al. 2007]{Liu07}). The total accretion mass is proportional to the accretion time in the model. Then the total energy radiated $E_{\rm iso}$ may be correlated with the duration $T_{\rm 90}$ (see also \cite[Gehrels et al. 2009]{Gehrels09}).

\section{Data and Implication}
{\underline{\it Data}}.
Data on fluence and $T_{\rm 90}$ are obained from:
$http://swift.gsfc.nasa.gov/docs/\\swift/archive/grb table/$ (Swift data) and $http://heasarc.gsfc.nasa.gov/$$W3Browse/\\fermi/fermigtrig.html$ (Fermi data)
up to September of 2012. GRBs with both short and long duration are included.
The left and middle panels of Fig. 1 show the correlation between the fluence and $T_{\rm 90}$.
As shown in the plot, the fluence is proportional to $T^{0.62\pm0.02}_{90}$
for the $Fermi$ data (left) and $T^{0.56\pm0.03}_{90}$ for the $Swift$ data (middle).
We further analyze the sources with known redshift in the $Swift$ sample.
For these GRBs, we find that ${E_{\rm iso}} \propto [T_{90}/(1+z)]^{1.01\pm0.15}$, as shown in the right panel of Fig. 1.
In addition, the short and long GRBs almost have the same trend
in the right panel except GRB 060218,
which is a low luminous GRB and may have different progenitor (\cite[Gehrels et al. 2009]{Gehrels09}).
\begin{figure}\label{Fig_1}
\centering\includegraphics[angle=0,scale=0.40]{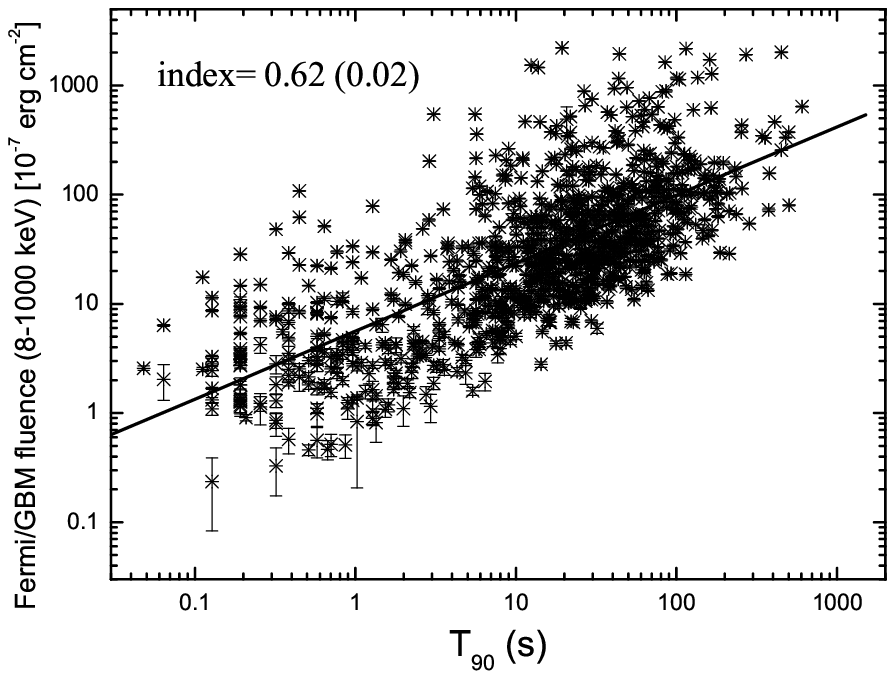}
\centering\includegraphics[angle=0,scale=0.40]{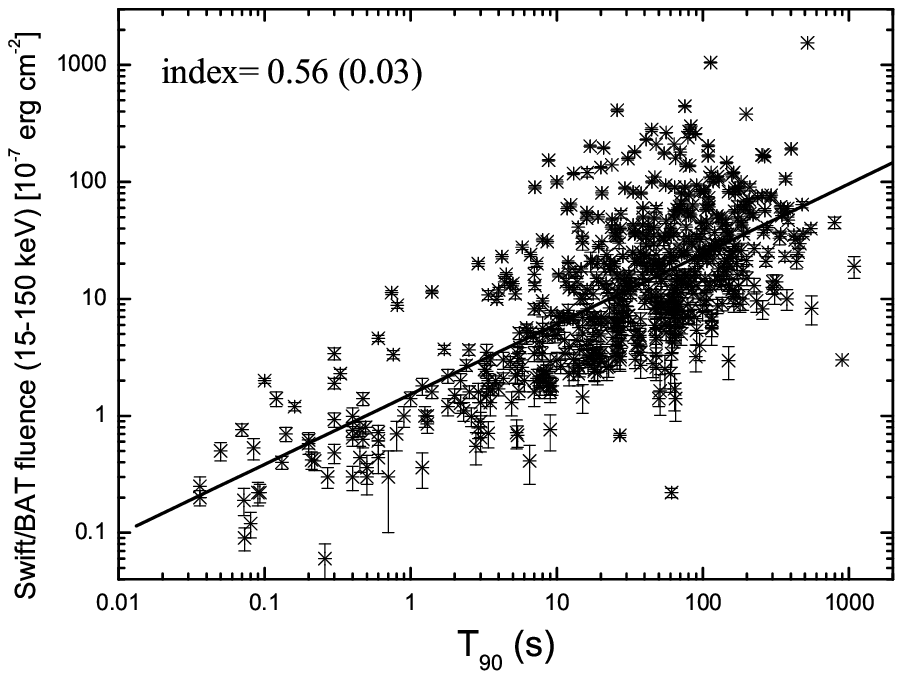}
\centering\includegraphics[angle=0,scale=0.40]{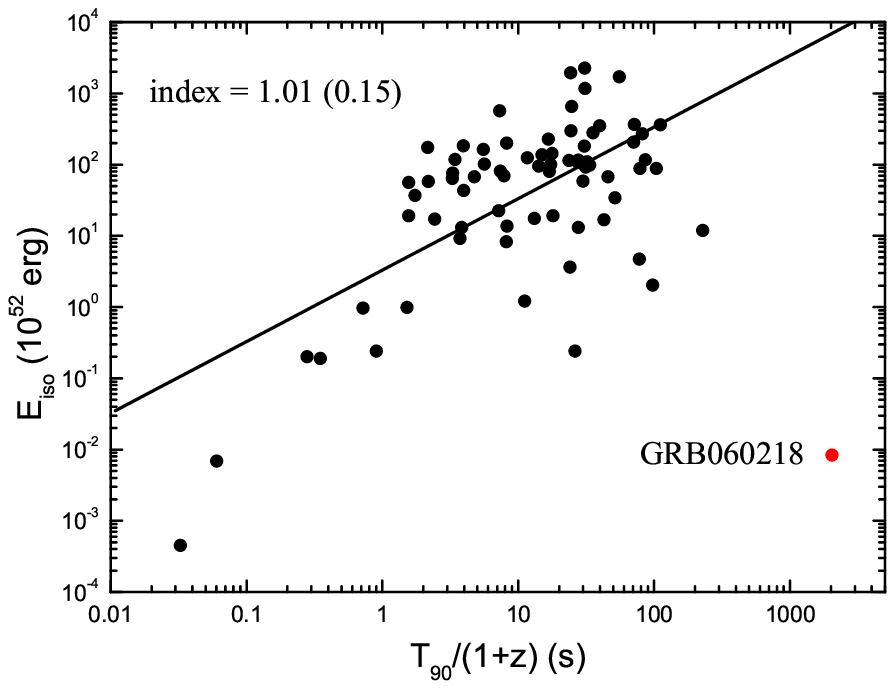}
\caption{Left and middle panel: Fluence $vs$ $T_{\rm 90}$. Right panel: $E_{\rm iso}$ $vs$ $T_{\rm 90}$ }
\end{figure}

{\underline{\it Implication}}.
In order to understand the correlation obtained in the above statistics,
we investigate the physical origin within the framework of the GRB-type accretion disk model.
Since $H\sim r$ in the advection-dominated accretion disk, we have
$t_{\rm vis} \sim \alpha ^{ - 1}\Omega _{\rm K}^{ - 1}$.
To simplify, we assume that the accreting mass $M_{\rm disk}$ is
deposited at the radius $r_{\rm disk}$.
Then, the timescale of the disk mass $M_{\rm disk}$ falling into the black hole
corresponds to the viscous timescale.
The equation can be described as (e.g., \cite[Narayan et al. 2001]{Narayan01})
\begin{equation}
{t_{\rm vis}} \sim 2.7{\alpha ^{ - 1}}\left(\frac{M_{\rm BH}}{M_{\odot}}\right)^{ - 1/2}\left(\frac{r_{\rm disk}}{10^9 \rm cm}\right)^{3/2}\;\;\rm s,
\end{equation}
\begin{equation}
{M_{\rm disk}} \sim 2\pi {r_{\rm disk}}\Sigma \Delta r.
\end{equation}
If the $r$-dependence disk surface density $\Sigma$ is $\Sigma  \propto {r^s}$ and $\Delta r\propto r$,
then we can get
\begin{equation}
{M_{\rm disk}} \propto t_{\rm vis}^{{{2(2 + s)} \over 3}}  \propto t_{\rm disk}^{{{2(2 + s)} \over 3}}.
\end{equation}
If we assume the $T_{\rm 90}$  equals the accretion time and the $E_{\rm iso}$ be proportional to the total mass of the disk,
we can derive,
\begin{equation}
{E_{\rm iso}} \propto T_{\rm 90}^{{{2(2 + s)} \over 3}}.
\end{equation}
Comparing the above equation with ${E_{\rm iso}} \propto T_{90}$, we get $s$ $\simeq$ -0.5.

\section{Conclusion and Discussion }
We consider that the disk of GRB is an advection-dominated disk in which the mass of the disk $M_{\rm disk}$ is deposited at the radius $r_{\rm disk} $. The accretion timescale of the disk into black hole corresponds to the viscous timescale. The correlation between $ {E_{\rm iso}}$ and ${{T_{\rm 90}}}$ obtained in statistics for GRBs can be qualitatively explained in the above model.

\acknowledgements
We acknowledge the use of the public data from the Swift and Fermi data archive. This work was supported by the National Basic Research Program of China (2009CB824800), the National Natural Science Foundation of China (11103015 and 11233006), and the Youth
Innovation Promotion Association of Chinese Academy of Sciences.

\end{document}